%% file: main.tex
\documentclass[sigconf]{acmart}

\usepackage{enumitem}
\usepackage{threeparttable}
\usepackage{multirow}
\usepackage{makecell}
\usepackage{pgfplots}
\usepackage{subcaption}
\usepackage{balance}
\usepackage{soul}
\usepackage[most]{tcolorbox}
\newtcolorbox{promptgreen}[1][]{
  colback=white,
  colframe=green!50!black,
  colbacktitle=green!50!black,
  coltitle=white,
  fonttitle=\bfseries,
  arc=2mm,
  title=#1
}
\newtcolorbox{promptblue}[1][]{
  colback=white,
  colframe=blue!50!black,
  colbacktitle=blue!50!black,
  coltitle=white,
  fonttitle=\bfseries,
  rounded  corners,
  title=#1
}
\newtcolorbox{promptlightgreen}[1][]{
  colback=white,        
  colframe=green!45!white,
  colbacktitle=green!45!white,
  coltitle=black,
  fonttitle=\bfseries,
  rounded corners,
  title=#1
}
\AtBeginDocument{%
  }

\copyrightyear{2026}
\acmYear{2026}
\setcopyright{cc}
\setcctype{by}
\acmConference[WWW '26]{Proceedings of the ACM Web Conference 2026}{April 13--17, 2026}{Dubai, United Arab Emirates}
\acmBooktitle{Proceedings of the ACM Web Conference 2026 (WWW '26), April 13--17, 2026, Dubai, United Arab Emirates}
\acmPrice{}
\acmDOI{10.1145/3774904.3792990}
\acmISBN{979-8-4007-2307-0/2026/04}

\begin{document}

\title{Cross-Domain Fake News Detection on Unseen Domains via LLM-Based Domain-Aware User Modeling}

\author{Xuankai Yang}
\affiliation{
  \department{School of Computing}
  \institution{Macquarie University}
  \city{Sydney}
  \country{Australia}}
\orcid{0000-0002-4985-3560}
\email{xuankai.yang@hdr.mq.edu.au}

\author{Yan Wang}
\authornote{Corresponding author.}
\affiliation{
  \department{School of Computing}
  \institution{Macquarie University}
  \city{Sydney}
  \country{Australia}}
\orcid{0000-0002-5344-1884}
\email{yan.wang@mq.edu.au}

\author{Jiajie Zhu}
\affiliation{
  \department{School of Computing}
  \institution{Macquarie University}
  \city{Sydney}
  \country{Australia}}
\orcid{0000-0001-8673-1477}
\email{jiajie.zhu@mq.edu.au}

\author{Pengfei Ding}
\affiliation{
  \department{School of Computing}
  \institution{Macquarie University}
  \city{Sydney}
  \country{Australia}}
\orcid{0000-0002-7048-7518}
\email{pengfei.ding@mq.edu.au}

\author{Hongyang Liu}
\affiliation{
  \department{School of Computing}
  \institution{Macquarie University}
  \city{Sydney}
  \country{Australia}}
\orcid{0000-0002-4201-1934}
\email{hongyang.liu2@hdr.mq.edu.au}

\author{Xiuzhen Zhang}
\affiliation{
  \department{School of Computing Technologies}
  \institution{RMIT University}
  \city{Melbourne}
  \country{Australia}}
\orcid{0000-0001-5558-3790}
\email{xiuzhen.zhang@rmit.edu.au}

\author{Huan Liu}
\affiliation{
  \department{School of Computing and Augmented Intelligence}
  \institution{Arizona State University}
  \city{Tempe}
  \country{USA}}
\orcid{0000-0002-3264-7904}
\email{huanliu@asu.edu}

\renewcommand{\shortauthors}{Xuankai Yang et al.}

\begin{abstract}
Cross-domain fake news detection (CD-FND) transfers knowledge from a source domain to a target domain and is crucial for real-world fake news mitigation. This task becomes particularly important yet more challenging when the target domain is previously unseen (e.g., the COVID-19 outbreak or the Russia-Ukraine war). However, existing CD-FND methods overlook such scenarios and consequently suffer from the following two key limitations: (1) insufficient modeling of high-level semantics in news and user engagements; and (2) scarcity of labeled data in unseen domains. Targeting these limitations, we find that large language models (LLMs) offer strong potential for CD-FND on unseen domains, yet their effective use remains non-trivial. Nevertheless, two key challenges arise: (1) how to capture high-level semantics from both news content and user engagements using LLMs; and (2) how to make LLM-generated features more reliable and transferable for CD-FND on unseen domains. To tackle these challenges, we propose \textbf{DAUD}, a novel LLM-based \underline{\textbf{D}}omain-\underline{\textbf{A}}ware framework for fake news detection on \underline{\textbf{U}}nseen \underline{\textbf{D}}omains. DAUD employs LLMs to extract high-level semantics from news content. It models users’ single- and cross-domain engagements to generate domain-aware behavioral representations. In addition, DAUD captures the relations between original data-driven features and LLM-derived features of news, users, and user engagements. This allows it to extract more reliable domain-shared representations that improve knowledge transfer to unseen domains. Extensive experiments on real-world datasets demonstrate that DAUD outperforms state-of-the-art baselines in both general and unseen-domain CD-FND settings.
\end{abstract}

\ccsdesc[500]{Information systems~Social networks}
\ccsdesc[500]{Computing methodologies~Neural networks}

\keywords{Cross-Domain Fake News Detection, Large Language Models, Social Media, User-News Engagement}


\maketitle

\input{sections/1introduction}
\input{sections/2relatedwork}
\input{sections/3method}
\input{sections/4experiments}
\input{sections/5conclusion}

\begin{acks}
This research is supported by ARC Discovery Projects DP200101441 and DP230100676, Australia.
\end{acks}

\bibliographystyle{ACM-Reference-Format}
\bibliography{mybase}
\appendix
\input{sections/6appendix}
\end{document}

%% file: sections/1introduction.tex
\section{Introduction}
\input{figures/intro_fig}

Fake news is widespread on social media, and its rapid dissemination has caused notable societal risks~\cite{grinberg2019fake, olan2024fake, rocha2021impact}, making fake news detection (FND) a crucial task~\cite{zhou2020survey}. Earlier FND studies focus on a single-domain setting, where they learn veracity-related features of news within a single domain. However, real-world fake news articles span diverse domains (e.g., politics, entertainment), which makes cross-domain fake news detection (CD-FND)~\cite{zhu2022memory, yue2023metaadapt, mosallanezhad2022domain, yang2024update} critical. This is because CD-FND methods can transfer knowledge from a source domain to improve the detection performance in a target domain. In addition, the CD-FND task becomes more significant yet more challenging when the target domain is previously unseen, arising from unexpected or rapidly unfolding events such as the COVID-19 outbreak~\cite{rocha2023impact} or the Russia–Ukraine war~\cite{khaldarova2020fake}. However, existing CD-FND studies have largely overlooked such scenarios.

Existing CD-FND studies mainly focus on extracting transferable linguistic and stylistic features from news content~\cite{wang2018eann, castelo2019topic, silva2021embracing, zhu2022memory, yue2023metaadapt}. In addition, some engagement-based studies explore cross-domain behavioral features based on user engagements (a.k.a., user interactions) in different domains~\cite{mosallanezhad2022domain, yang2024update, yang2025macro}. Despite the efforts, these two types of methods are unable to effectively handle news from previously unseen domains. Consequently, existing CD-FND approaches face two fundamental limitations (LMs).

\textbf{LM1: }\textit{Existing CD-FND methods primarily rely on surface-level semantics from news content and user engagements, overlooking high-level semantics}. As a result, these methods fail to develop a deeper understanding of news content and user engagements, which thereby undermines CD-FND performance in a general CD-FND setting. For example, as illustrated in Fig.~\ref{fig:intro_fig} (a), a journalist user reposted a political and an entertainment news articles, both of which relate to the theme of key-person changes. However, existing CD-FND methods overlook such high-level semantics (e.g., both concerning key-person changes) and instead rely on surface-level semantics (e.g., both happening in New York). Such process results in user features that fail to capture the journalist user’s actual news preferences for major news events, particularly key-person changes. These inaccurate features ultimately degrade detection performance, especially in unseen domains where stable semantic features are crucial for generalization.

\textbf{LM2: }\textit{Existing CD-FND methods can hardly extract effective features from news content and user engagements in unseen domains with limited labels}. In practice, new domains continually emerge with considerably different linguistic and behavioral characteristics. However, existing methods cannot be applied directly to unseen domains because no labeled data are available for adaptation. For example, as illustrated in Fig.~\ref{fig:intro_fig} (b), during the COVID-19 outbreak, a large volume of news emerged within a short period, consequently existing methods do not generalize well to such an unseen domain. 

These limitations motivate the need for a more adaptable approach to capture high-level semantics from news content and user engagements. Recently, large language models (LLMs) have demonstrated strong capabilities in several aspects relevant to unseen-domain CD-FND. Some single-domain FND methods~\cite{hu2024bad, wang2024explainable, hang2024trumorgpt, ma2024fake} leverage LLMs to capture rich semantic information from news content even when annotated labels are limited. However, effectively leveraging LLMs to this unseen-domain CD-FND setting remains non-trivial due to the following two challenges (CHs):

\textbf{CH1}: \textit{How to capture high-level semantics from both news content and user engagements using LLMs?} As mentioned earlier, LLMs can extract high-level semantics from news content. Nevertheless, the high-level semantics contained in user engagements are implicit, which often leads LLMs to infer inaccurate or biased user preferences. In practice, existing LLM-based FND methods~\cite{wan2024dell, nan2024let} attempt to model users behaviors by prompting LLMs with selected user attributes (e.g., gender, age, or occupation). These approaches rigidly assume that users sharing the same attributes exhibit similar behavioral preferences, failing to capture the implicit high-level semantics encoded in actual user engagements, thus limiting their effectiveness in CD-FND.

\textbf{CH2}: \textit{How to make LLM-generated features more reliable and transferable for CD-FND on unseen domains?} Although LLMs can extract features from unseen-domain news content and user engagements without labeled data, these features may introduce hallucinated information when not grounded in real data. For example, an LLM may infer facts that do not appear in the original text, such as making up explanations or adding invented participants. Such unreliable outputs introduce noise into the transferred knowledge and limit generalization to the unseen domain.

\textbf{Our Approach and Contributions.} To address the above two challenges, in this paper, we propose a novel LLM-based \underline{\textbf{D}}omain-\underline{\textbf{A}}ware framework for CD-FND on \underline{\textbf{U}}nseen \underline{\textbf{D}}omains, named \textbf{DAUD}. To the best of our knowledge, \ul{this is the first work} to propose a CD-FND framework that operates under an unseen-domain setting. Our framework consists of an \underline{L}LM-based \underline{D}omain-\underline{A}ware \underline{E}nhancement module (LDAE) and a \underline{D}omain-\underline{S}hared feature learning and \underline{R}elation-aware \underline{A}lignment module (DSRA). The contributions of this work are summarized as follows:
\begin{itemize}[leftmargin=*]
    \item Targeting \textbf{CH1}, in LDAE module, 
    we design an LLM-based domain-aware user agent to learn high-level behavioral features for user from their single- and cross-domain engagements. Based on such user agent, our LDAE module extracts high-level semantics from news content, and augments user engagements based on users' high-level behavioral preferences.
    \item Targeting \textbf{CH2}, in DSRA module, we design an approach that captures the relations between the original data-driven features and the LDAE-generated features of news, users and user-news engagements, and further learns domain-shared features from these three corresponding information sources in different domains. This facilitates more reliable and effective cross-domain knowledge transfer, thereby mitigating the problem of insufficient domain information in unseen domains.
    \item 
    Extensive experiments on three real-world datasets demonstrate that our proposed DAUD framework significantly outperforms state-of-the-art baselines across all evaluation metrics, confirming its improved performance in both general and unseen-domain CD-FND settings.
\end{itemize} 

%% file: figures/intro_fig.tex
\begin{figure}
  \centering
  \includegraphics[scale=0.52]{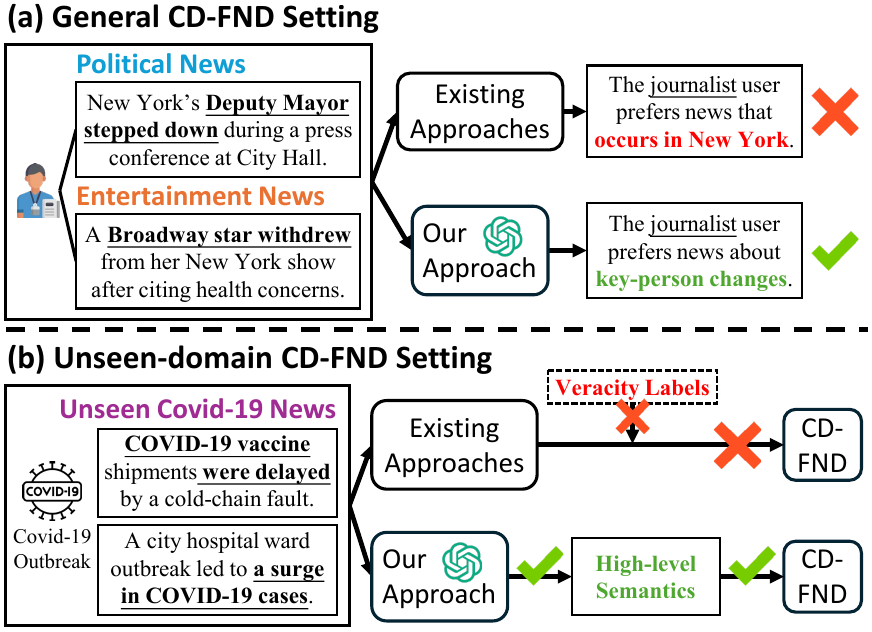}
  \caption{Our LLM-based approach provides advantages over existing approaches in both general and unseen-domain CD-FND settings.}
  \label{fig:intro_fig}
  \Description{illustrate of our task}
\end{figure}

%% file: sections/2relatedwork.tex
\section{Related Work}
\subsection{Single-Domain Fake News Detection}
\subsubsection{Traditional Single-Domain Methods}
Fake news detection can be categorized into news content-based and user engagement-based methods according to the data they rely on.
Content-based studies have leveraged diverse types of features for FND, including lexical features~\cite{perez2017automatic, volkova2017separating}, sentence-level and contextual-level representations~\cite{ma2016detecting, shu2019defend}.
Nevertheless, these methods depend heavily on news content itself and ignore user engagements signals, making them sensitive to textual differences in news and thereby limiting FND performance.
To more effectively utilize user engagement information, existing studies incorporate social context signals into the detection process. For example, some studies enhance fake news detection by incorporating user engagement information as additional social signals~\cite{nguyen2020fang, khoo2020interpretable}. And studies in~\cite{zhu2024propagation} further utilities engagement propagation patterns to capture broader diffusion signals.
However, due to limited reasoning and deep understanding ability in their underlying models, these methods cannot capture high-level semantics.

\subsubsection{LLM-based Single-Domain Methods}
The rapid development of LLMs has motivated many single-domain FND methods that leverage these models to better interpret news content and user engagements.
Some news content-based studies utilize LLMs to generate multi-perspective rationales or justifications for veracity reasoning~\cite{hu2024bad, wang2024explainable}, while others leverage LLMs to obtain factual evidence through retrieval-augmented generation or to extract higher-level entity–topic semantics~\cite{hang2024trumorgpt, ma2024fake}.
The user engagement-based methods prompt LLMs to derive social signals from readership patterns for aligning few-shot predictions~\cite{wu2023prompt}, while others generate LLM-based reactions or comments to emulate diverse user behaviors and expand engagement contexts~\cite{wan2024dell, nan2024let}.
Overall, these above-mentioned approaches rely on data from only a single domain, preventing them from leveraging complementary information across domains that could improve detection performance.

\subsection{Cross-Domain Fake News Detection}
\subsubsection{Traditional Cross-Domain Methods}
Recent studies have begun to explore CD-FND, which aims to learn representations that remain effective across different news domains.
To this end, diverse studies aim to extract domain-shared features from news content~\cite{castelo2019topic, zhu2022memory, yue2022contrastive, wang2018eann, silva2021embracing, yue2023metaadapt}.
Cross-domain studies also incorporate user engagement information to facilitate knowledge transfer. Some works leverage cross-domain engagements to derive domain-shared representations~\cite{mosallanezhad2022domain, yang2024update}, while others integrate both micro-level content representation and macro-level engagement behaviors to support transfer across domains~\cite{yang2025macro}.
In addition, some unsupervised approaches~\cite{ran2023unsupervised, yue2022contrastive} detect fake news without target-domain labels by reducing the gap in feature distributions across domains. However, lacking explicit control over the transferred information inadvertently leads to unintended noise.

\subsubsection{LLM-based Cross-Domain Methods}
The enhanced reasoning and comprehension abilities of LLMs have increasingly motivated their application in CD-FND.
Building on this trend, recent studies explore LLMs for CD-FND from several practical angles. Some studies employ LLMs to facilitate multi-agent analysis or decision-making processes for target-domain news~\cite{li2025multi}, whereas others use affect-aware retrieval to assemble cross-domain examples for in-context reasoning~\cite{liu2025raemollm}.
In addition, several approaches utilize LLMs to assist model adaptation under distribution shift to improve rumor or misinformation detection~\cite{gong2025cross, liu2025montrose}.
However, these approaches do not explicitly transfer knowledge across domains, which leaves them unable to capture the stable veracity signals that are essential for reliable detection in the target domain.
In contrast, our approach specifically models cross-domain user behaviors by combining LLM-derived semantics from both news content and user engagements. Furthermore, our approach aligns them through hierarchical contextual relations to enhance detection performance in an unseen-domain CD-FND setting.

%% file: sections/3method.tex
\section{Methodology}
\input{figures/general_fig}
\input{figures/overall_fig}
\input{figures/agent_fig}
\subsection{Problem Statement}
Let $\mathcal{G}$ denote the set of news domains. For each domain $g \in \mathcal{G}$, the labeled dataset is $\mathcal{D}_g = \{ (x_i, y_i) \}_{i=1}^{N_g}$, where $N_g$ is the number of news items in domain $g$. Here, $x_i$ is the $i$-th news item, and $y_i$ is its veracity label indicating whether the news is fake or true. Each news item $x_i$ is associated with a set of user engagements $\mathcal{E}_i = \{(u_j, c_j)\}_{j=1}^{M_i}$, where $u_j$ is a user, $c_j$ is the user comment, and $M_i$ is the total number of users who engaged with $x_i$. 
Users often engage with news across different domains. Let $\mathcal{U}_S$ denote the user set of the source domains and $\mathcal{U}_t$ the user set of the target domain. Users in the intersection $\mathcal{U}_S \cap \mathcal{U}_t$ are common users, whose engagements across domains provide transferable behavioral patterns.

With these definitions, we now consider the \textbf{unseen-domain CD-FND setting}. Given source-domain labeled data $\{\,\mathcal{D}_g\,\}_{g \neq t}$ and no labeled data from the target domain $\mathcal{D}_t$, the goal is to predict the veracity label $y_i$ of each target-domain news item $x_i$ by leveraging (1) its content, (2) its associated engagements $\mathcal{E}_i$, and (3) cross-domain behavioral information provided by common users.

\subsection{Model Overview}
The learning process of DAUD is illustrated in Fig.~\ref{fig:general}.
DAUD contains two key components: (1) LLM-based Domain-Aware Enhancement module (\textit{LDAE}), 
which extracts high-level semantics with LLMs to enrich news and user features and to augment user engagements across domains;
(2) Domain-Shared Feature Learning and Relation-aware Alignment (\textit{DSRA}), which removes domain-specific and unreliable information from both original data-driven and LLM-generated features, and extracts stable domain-shared representations based on these refined features.
DSRA is trained on two source domains to learn transferable representations and is directly applied to the unseen target domain. Together, LDAE and DSRA enable DAUD to perform CD-FND on unseen domains.

\subsection{LLM-based Domain-Aware Enhancement (LDAE)}
As illustrated in Figure~\ref{fig:overall}, LDAE operates on three levels: (1) news feature enrichment, (2) user behavior modeling, and (3) personalized engagement augmentation. These components collectively provide semantically enriched inputs for downstream domain-shared representation learning.
\subsubsection{News Feature Enrichment}
Given the content of a news article $x_i$, we design a prompt-driven procedure, named News Feature Enrichment ($\operatorname{NFE}$), to obtain an enriched news summary $d_i$ that captures semantics relevant to news veracity. The news feature extraction prompt guides the LLM to highlight semantic characteristics that help distinguish fake news, including even highly camouflaged cases. 
To ensure that these extracted features are aligned with the needs of fake news detection, the prompt decomposes the analysis into several interpretable steps that cover domain style, sentiment, structural feature, logical consistency, and source credibility. These dimensions correspond to common features used by human fact-checkers, enabling the LLM to focus on veracity-related features.
For brevity, the news feature extraction prompt is included in Appendix~\ref{app:prompts}.

\subsubsection{User Behavior Modeling}
To obtain a richer and more reliable representation of each user, we further collect their historical engagements on news items other than the target news. For user $u_j$, the historical engagements is $\{\, (e_k, c_k) \,\}_{k=1}^{K_j}$, where $e_k$ is a historical news item, $c_k$ is the corresponding user comment, and $K_j$ is the total number of historical engagement records.
We design a domain-aware user agent that derives two complementary user representations: a domain-aware user profile $p_j$ based on news content the user has engaged with, and a user commenting feature $s_j$ derived from the user’s historical comments. 
For user engagements, we represent each engaged news item using the summary extracted by our $\operatorname{NFE}$, and feed the news summary into the domain-aware user agent. For comments, we design a commenting feature extraction prompt that instructs the LLM to infer the user’s commenting style reflected in their historical comments. Together, these two features model users from both their engagement and commenting behaviors, providing a more complete characterization.

The overall structure of the user agent is illustrated in Fig. \ref{fig:user_agent}. The process of generating domain-aware user profiles consists of two steps. 
In Step 1, an engagement prediction prompt is constructed using the generated user profile together with the summaries of the historical news items and news content of them. The prompt instructs the LLM to predict whether the user would choose to engage with each news item. The engagement prediction prompt is presented below.

\begin{promptgreen}[1. Engagement Prediction Prompt]
\textbf{You are simulating the behavior of a Twitter (X) user, who is either a regular user, a debunking user, or a malicious user (e.g., spammer or troll).}

Here is your self-introduction about what type of user you are as well as the types and characteristics of news you like or dislike engaging with: \{\,\textcolor{blue}{\textit{user profile}}\,\}. 
You are now evaluating whether to repost the following news article based on this user’s perspective: \{\,\textcolor{blue}{\textit{news article}}\,\}; Its news features are: \{\,\textcolor{blue}{\textit{news features}}\,\}.

Follow these steps:

(1) Review your self-introduction to \textbf{identify the user type, news types you prefer and dislike}; 

(2) Analyze the features of the news article (e.g., \textbf{News Domain; Sentiment; Structural Features; Logical Consistency; and Source Credibility}); 

(3) Assess whether these features align or conflict with your preferences; 

(4) Decide whether to Repost or Ignore the news. Provide a \textbf{detailed explanation} that connects your decision to your self-introduction and the news features.
\end{promptgreen}

In Step 2, the predicted engagements are compared against the user’s historical engagements. 
A user features update prompt guides the LLM to revise the user profile to better reflect the user’s actual engagements.
This refinement is performed iteratively until the predicted engagements become consistent with the actual history, yielding a domain-aware user profile. The user features update prompt is shown below.

\begin{promptgreen}[2(a). User Features Update Prompt] 
Here is your current self-introduction describing your user type and the types and characteristics of news you like or dislike engaging with: \{\,\textcolor{blue}{\textit{user profile}}\,\}. Recently, you predicted that this user would ignore the following news article: \{\,\textcolor{blue}{\textit{news article}}\,\}; Its news features are: \{\,\textcolor{blue}{\textit{news features}}\,\}; Your explanation was: \{\,\textcolor{blue}{\textit{user explanation}}\,\}.

However, the user actually reposted the news article. 
\end{promptgreen}

\begin{promptgreen}[2(b). User Features Update Prompt]
This indicates that your self-introduction may be inaccurate, incomplete, or missing a key motivational factor that caused this action. 
\textbf{Your task is to revise your self-introduction so that it can explain the reposting behavior naturally and accurately.}
\textbf{Follow these steps:}

(1) Identify what you overlooked or misunderstood about the news article that led to the repost; 

(2) Analyze what features of the article (i.e., news domain; sentiment; structural features; logical consistency; and source credibility) may motivate the user to repost it; 

(3) Consider how your user type or value system may need to be updated to reflect this motivation. 

(4) Decide which past preferences should be retained, revised, or discarded to avoid future contradiction; 

(5) Write an updated self-introduction that: \textbf{Starts with your new user type; Describes your newfound preferences reflected in this interaction; Summarizes any relevant retained preferences; Describes what types of news you now dislike}.
\end{promptgreen}

After the above two steps, the final user profile captures key behavioral characteristics such as user type and preferences. Since the profile is learned through iterative refinement grounded in real engagement behaviors rather than static attributes, it provides a more reliable profile of the user and serves as a solid basis for subsequent feature enrichment and personalized engagement augmentation.

\subsubsection{Personalized Engagement Augmentation}
Based on the news feature enrichment and user behavior modeling components in LDAE, we obtain a more accurate domain-aware user profile $p_j$ and user commenting feature $s_j$ for each user. While the user profile encodes the user’s preferences, converting these preferences into additional engagement instances provides more explicit behavioral features. Therefore, we design a personalized engagement augmentation consisting of three steps, which are detailed below.

Firstly, we identify news articles that user $u_j$ has not engaged with, and then select from those that are most similar to the user’s historical engagement items. We denote the resulting unengaged news set as $A_j = \{\, a_t \,\}_{t=1}^{T_j}$, where $T_j$ is the number of selected unengaged news items for user $u_j$.
Secondly, we apply the engagement prediction prompt from the user agent to determine which of these unengaged news items the user would choose to engage with. 
We denote the set of items predicted to be engaging as $\hat{A}_j \subseteq A_j$.
Thirdly, we use a comment generation prompt to derive personalized commenting features for the news items predicted as engaging. The prompt takes the user profile $p_j$, the user’s commenting features $s_j$, and the predicted engagements set $\hat{A}_j$ as input, and generates comments that reflect the user’s writing style and preferences. We denote the resulting set of generated comments as $V_j = \{\, v_t \,\}_{t=1}^{|\hat{A}_j|}$. For completeness, the full comment-generation prompt is provided in Appendix~\ref{app:prompts}.

By predicting user-news engagements and generating comments based on a domain-aware user profile, the personalized engagement augmentation provides richer and more realistic behavioral information. This results in enhanced and more informative inputs for subsequent domain-shared feature learning.

\subsection{Domain-Shared Feature Learning and Relation-Aware Alignment (DSRA)}
DSRA module is designed to obtain more effective domain-shared representations for CD-FND, particularly in the unseen-domain setting. DSRA achieves this goal through two key components. 
Firstly, DSRA mitigates noise and domain-specific biases that within both original data-driven and LLM-generated features of news, users and user engagements. Secondly, DSRA extracts the relations between these two types features and then aligns the resulting representations to obtain fine-grained domain-shared ones.

\subsubsection{Multi-Level Domain-Shared Feature Learning}
To effectively learn transferable representations for CD-FND, DSRA performs multi-level domain-shared feature learning across news, user and user engagement.
We begin by applying a hierarchical disentangler to each type of data. Inspired by~\cite{yang2025macro}, this disentangler first extracts veracity-relevant features and then derives domain-shared components from them. This step is applied to all three data types before detailing each level below.
\textbf{News level:} for each news article $x_i$, we use two features: the content embedding $h_i^{x}$ from the raw text and the summary embedding $h_i^{d}$ generated by the $\operatorname{NFE}$. Both features contain domain-specific information. The disentangler produces their domain-shared representations, denoted as $z_i^{x}$ and $z_i^{d}$.
\textbf{User level:} for each user $j$, we use the domain-aware user profile generated by the LLM-based user agent. Its embedding $h_j^{p}$ captures the user’s behavioral tendencies across domains, yet still contains domain-specific signals. The disentangler extracts the corresponding domain-shared representation, denoted as $z_j^{p}$.
\textbf{Engagement level:} for each engagement $k$, we consider two types of features: 
(1) the embedding of user engagements $h_k^{e}$ (including both original and LLM-augmented engagements), and (2) the embedding of user comments $h_k^{c}$ (including the original comments as well as LLM-generated comment augmentations). The engagements and comments reflect domain-specific behavioral preferences and commenting styles. Applying the disentangler extract their domain-shared representations, denoted as $z_k^{e}$ and $z_k^{c}$, respectively.

\subsubsection{Relation-Aware Alignment}
In this section, we address two key objectives. Firstly, we individually extract relational signals within each level of news, user and user engagements, thereby avoiding inaccuracies caused by hallucination.
Secondly, we align the representations from news, users, and user engagements. This alignment provides complementary signals derived from users and their engagements for fake news detection.

For the first objective, DSRA incorporates three fusion modules to extract relational signals within each level and enhance the corresponding representations. 
Let $\mathbf{z}_i^{x}$ and $\mathbf{z}_i^{d}$ denote the domain-shared representations of the news content and news summary, respectively. In Mutual Relation Fusion $\operatorname{MRF}$, we first perform relation modeling to extract the interaction between the news content and its LLM-generated summary:
\begin{equation}
\begin{aligned}
\mathbf{r}_i &= \operatorname{Rel}\!\left(\mathbf{z}_i^{x}, \mathbf{z}_i^{d}\right).
\end{aligned}
\end{equation}
This relation vector is then injected back into both sides to obtain relation-enhanced representations, which are fed into a bi-directional co-attention operator:
\begin{equation}
\hat{\mathbf{z}}_i^{x},\, \hat{\mathbf{z}}_i^{d}
= \operatorname{CoAtt}\!\left(
\mathbf{z}_i^{x} + \mathbf{W}_{x}\mathbf{r}_i,\;
\mathbf{z}_i^{d} + \mathbf{W}_{d}\mathbf{r}_i
\right).
\end{equation}
Finally, the mutually attended features are concatenated to obtain the enhanced news representation $\mathbf{z}_i^{F} = [\,\hat{\mathbf{z}}_i^{x} \,\|\, \hat{\mathbf{z}}_i^{d}\,]$.

Profile-aware Fusion leverages user profiles as references to enhance user representations. Given the domain-shared user profile sequence $\mathbf{Z}_j^{p}$ and the user behavior encoding sequence $\mathbf{Z}_j^{E}$ extracted from user engagements, the module first computes relation vectors $\mathbf{R}_j$. This relation is then injected into the behavior encoding sequence and refined through a cross-attention operator that uses user profiles as the reference: 
\begin{equation}
\hat{\mathbf{Z}}_j^{E}
= \operatorname{CrossAtt}\!\left(
\mathbf{Z}_j^{E} + \mathbf{W}_{E}\mathbf{R}_j,\;
\mathbf{Z}_j^{p}
\right).
\end{equation}

Comment-aware Fusion operates in a similar way, using the augmented comment sequence $\mathbf{Z}_j^{c}$ as a semantic reference to enhance the engagement representations $\mathbf{Z}_j^{e}$. It derives relation vectors between engagements and comments, injects it into the engagement encoding sequence, and applies cross-attention with comments as the reference to obtain the refined engagement representation $\hat{\mathbf{Z}}_j^{e}$.

Following the fusion operations in each level, DSRA performs a sequential alignment from the user engagement level to the user level and finally to the news level. 
\textbf{Engagement level.} After Comment-aware Fusion, we obtain the refined engagement encoding sequence $\hat{\mathbf{Z}}_j^{e}$. A Transformer encoder is applied to this sequence to extract a user behavior representation: $\mathbf{z}_j^{E} = \operatorname{TransEnc}(\hat{\mathbf{Z}}_j^{e})$, where $\mathbf{Z}_i^{E}$ denotes the user behavior encoding sequence, and each user behavior encoding is obtained by aggregating all engagements of user $j$.
\textbf{User level.} After Profile-aware Fusion, we similarly encode the sequence of domain-shared user representations using a Transformer to obtain a user pattern encoding tailored to the target news, i.e., $\mathbf{z}_i^{U} = \operatorname{TransEnc}(\hat{\mathbf{Z}}_i^{E})$.
\textbf{News level.} Finally, DSRA aligns the user pattern encoding $\mathbf{z}_i^{U}$ with the news representation produced by Mutual Relation Fusion $\mathbf{z}_i^{N}$. These two aligned signals together form the input to the final fake news detection module.

\subsection{Fake News Detection}
Given the final news representation $\mathbf{z}_i^{N}$ and the user pattern encoding $\mathbf{z}_i^{U}$, DAUD concatenates them and feeds the combined representation into a prediction layer to estimate the veracity label:
\begin{equation}
\hat{y}_i = \sigma\!\left(\mathbf{W}[\mathbf{z}_i^{N} \Vert \mathbf{z}_i^{U}] + b\right).
\end{equation}
The model is trained using the binary cross-entropy loss:
\begin{equation}
\mathcal{L}_{\text{det}} = - \left( y_i \log \hat{y}_i + (1-y_i)\log(1-\hat{y}_i) \right).
\end{equation}

%% file: figures/general_fig.tex
\begin{figure}[t]
  \centering
  \includegraphics[scale=0.47]{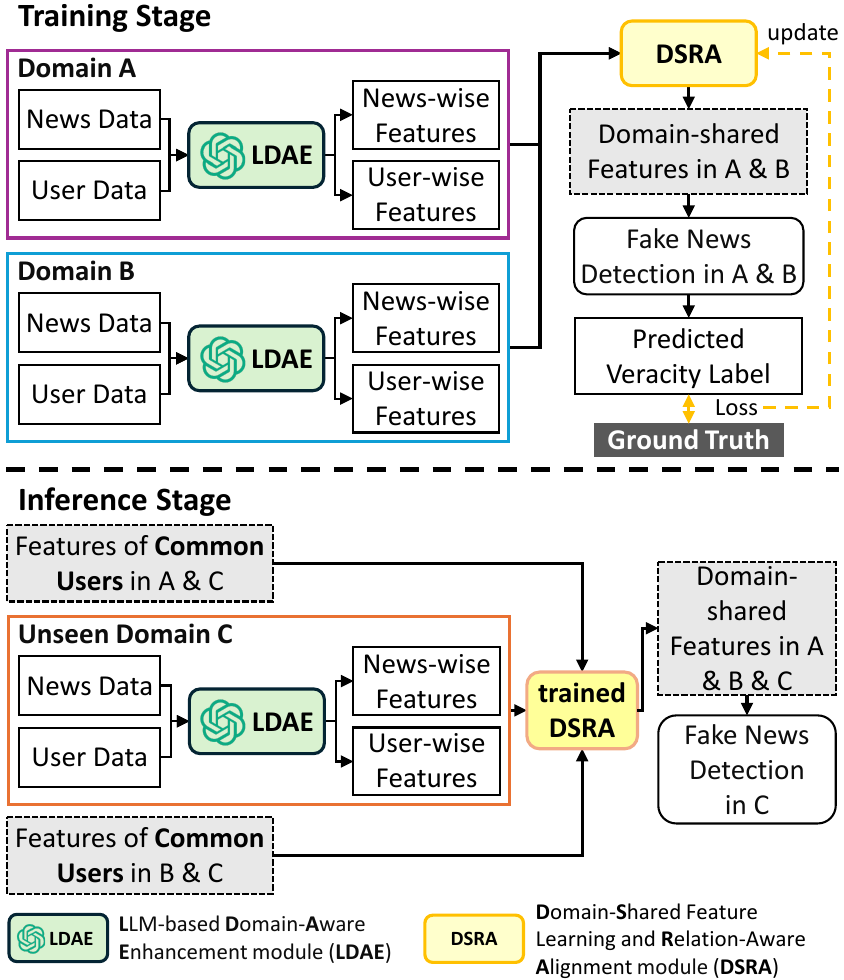}
  \caption{The learning process of DAUD framework.}
  \label{fig:general}
  \Description{general framework}
\end{figure}

%% file: figures/overall_fig.tex
\begin{figure*}[t]
  \centering
  \includegraphics[scale=0.6]{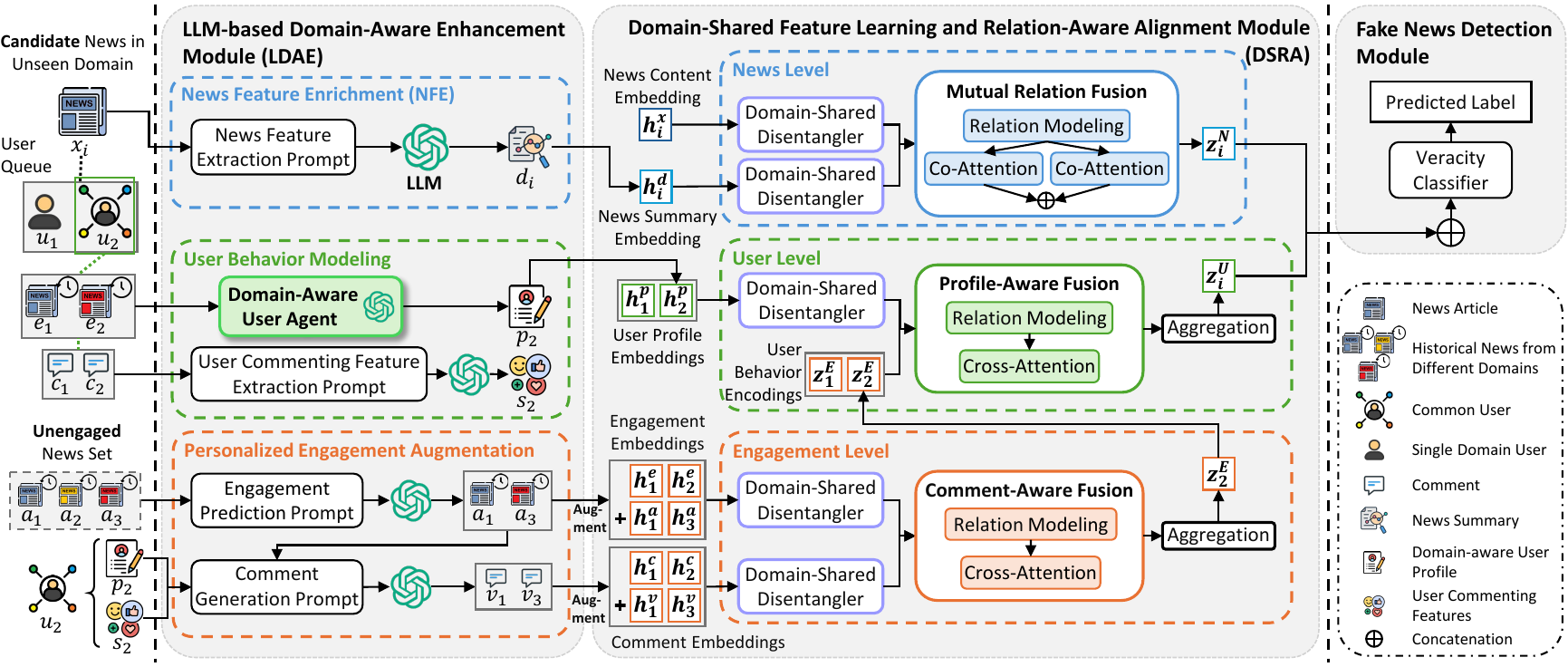}
  \caption{The overall architecture of DAUD framework.}
  \label{fig:overall}
  \Description{overall framework}
\end{figure*}

%% file: figures/agent_fig.tex
\begin{figure}
  \centering
  \includegraphics[scale=0.68]{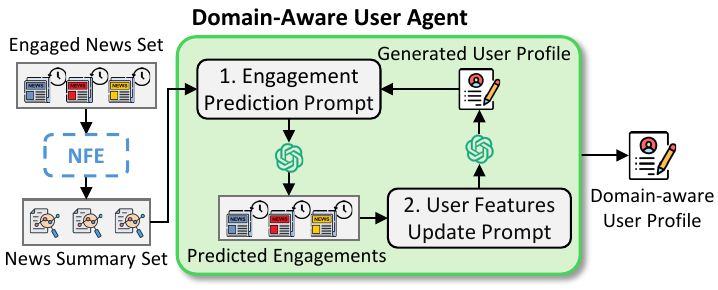}
  \caption{The architecture of Domain-Aware User Agent.}
  \label{fig:user_agent}
  \Description{domain-aware user agent}
\end{figure}

%% file: sections/4experiments.tex
\section{Experiments and Analysis}
\input{tables/general_performance}
We conduct extensive experiments to answer the following research questions (RQs): 
\begin{description}[leftmargin=*]
    \item[RQ1.] How does DAUD perform compared to state-of-the art baselines in general and unseen-domain CD-FND settings?
    \item[RQ2.] How does each proposed module in DAUD contribute to the overall CD-FND performance?
    \item[RQ3.] Can our proposed domain-aware user agent generate effective user profiles and user comments?
\end{description}

\subsection{Experimental Setup}
\subsubsection{Datasets}
We use data from three widely studied news domains: \emph{Politics (Polit.)}~\cite{shu2020fakenewsnet}, \emph{Entertainment (Enter.)}~\cite{shu2020fakenewsnet}, and \emph{COVID-19 (Covid.)}~\cite{li2020mmcovid}.
It is worth mentioning that news in different domains substantially differ in both content and linguistic characteristics. The detailed illustration is provided in Appendix~\ref{app:datasets}.
\subsubsection{Fake News Detection Settings}
We consider the following two distinct scenarios:
\textbf{(1) Unseen-domain CD-FND setting.}
In this setting, two of the three domains are used as \textit{source domains} and the remaining one domain serves as the unseen \textit{target domain}. We repeat this procedure three times so that each domain serves once as the unseen target domain. 
In addition to the characteristic differences between the source and target domains, it is also worth mentioning that no data from the unseen target domains is used during training on the source domains, ensuring our evaluation is strictly within the unseen-domain setting.
\textbf{(2) General CD-FND setting.}
In this setting, we use the similar procedure as in the unseen-domain setting: two domains are used as sources and the third one as the target, and this process is repeated until each domain has served once as the target.
The key difference from the unseen-domain setting is that the target domain’s training data is used during training of source domains.

\subsubsection{Baselines}
To comprehensively evaluate our framework, we compare DAUD with thirteen baseline methods from three categories: 
\textbf{(1) single-domain traditional methods} include BERT~\cite{devlin2018bert}, dEFEND~\cite{shu2019defend}, FANG~\cite{nguyen2020fang}, and DUCK~\cite{tian2022duck}, (2) \textbf{single-domain LLM-based methods} include GPT-4o-mini, ARG~\cite{hu2024bad}, and DELL~\cite{wan2024dell}, and (3) \textbf{cross-domain traditional methods} include EANN~\cite{wang2018eann}, Real-FND~\cite{mosallanezhad2022domain}, EDD~\cite{silva2021embracing}, M3FEND~\cite{zhu2022memory}, UPDATE~\cite{yang2024update} as well as MMHT~\cite{yang2025macro}. 
Note that we do not include the above-mentioned LLM-based CD-FND methods as baselines because they do not explicitly perform knowledge transfer, making them incompatible with our evaluation settings.
Detailed descriptions of these baseline methods are provided in Appendix~\ref{app:baseline}.
In the unseen-domain setting, we compare DAUD with cross-domain traditional baselines, because single-domain baselines cannot transfer knowledge across domains.
In the general CD-FND setting, we compare DAUD with cross-domain traditional baselines and all single-domain baselines. 
Note that for the single-domain methods, we train them on each domain and report their performance.

\subsubsection{Implementation Details}
Models are trained with Adam and evaluated using Acc, Prec, Rec, F1, and AUC, averaged over five runs. More implementation details are provided in Appendix~\ref{app:implement}.

\subsection{Main Results}
\input{tables/unseen_performance}

\subsubsection{General and Unseen-Domain CD-FND Performance Comparison (RQ1)}
Tables~\ref{tab:general} and~\ref{tab:unseen} present the performance comparison between DAUD and baselines in both general and unseen-domain CD-FND settings. 
Overall, DAUD achieves consistent and notable improvements over the best-performing baselines across all settings.
Specifically, in the general CD-FND setting, DAUD achieves average improvements of 1.22\% in Politics, 2.50\% in Entertainment, and 3.15\% in COVID-19. In the unseen-domain CD-FND setting, the improvements become more pronounced, reaching 1.96\%, 2.43\%, and 3.62\% in the three domains, respectively.
These improvements come from the complementary roles of LDAE and DSRA. LDAE's fine-grained semantic extraction enriches the learned representations, while DSRA enhances cross-domain knowledge transfer with learning stable domain-shared features. In contrast, existing CD-FND methods either overlook high-level semantic relations across news, users, and engagements, or depend heavily on domain-specific information. Such limitations prevent them from learning reliable domain-shared representations, leading to degraded performance in unseen domains where linguistic characteristics and user behavior patterns differ and labeled data are limited.

Furthermore, it is worth noting that in Tables~\ref{tab:general} and~\ref{tab:unseen}, almost all baseline methods experience performance drops when evaluated on unseen domains. Across the unseen-domain evaluations, UPDATE , M3FEND  and EDD  suffer the largest declines with average AUC declines of 32.27\%, 28.76\%, and 27.10\% respectively. This is because they depend heavily on domain-specific features, which fail to generalize to unseen domains.

\subsubsection{Ablation Study (RQ2)}
\input{figures/ablation_study}
To evaluate the contributions of the core modules in DAUD, we construct two variants: (1) \textbf{w/o LDAE}, which removes fine-grained semantic enrichment and augmentation, and (2) \textbf{w/o DSRA}, which removes relation-aware alignment across news, users, and engagements.
Ablation results in the unseen-domain CD-FND setting are shown in Fig.~\ref{fig:ablation}.\footnote{The general CD-FND setting exhibits the same performance trends.}
Across all three domains, w/o LDAE shows consistent performance drops: 3.90\% on Politics, 4.19\% on Entertainment, and 7.22\% on COVID-19, demonstrating that LDAE is essential for generating enriched semantic features and augmenting behavioral information.
The performance degradation becomes more substantial in w/o DSRA, 
with declines of 10.32\%, 9.32\%, and 8.76\% in the three domains. This confirms that DSRA’s relation-aware alignment between original data-driven and LLM-derived features is crucial for learning stable domain-shared representations.
Overall, these results show that both LDAE and DSRA are indispensable.

\subsubsection{Quality of Generated User Profiles and Augmented Engagements (RQ3)}
To evaluate the quality of the generated user profiles and the augmented engagements, we perform two complementary analyses.
Firstly, we assess the quality of the comments generated by LDAE by comparing them with real comments across all domains. The results show that the generated comments achieve higher overall quality and more stable performance than real comments. Detailed evaluation procedures and results are reported in Appendix~\ref{app:rq3}.

Secondly, we examine whether the learned user profiles capture meaningful user preferences by comparing them with predefined user attributes in an auxiliary engagement prediction task. The results show that LDAE substantially outperforms the attribute-based baseline, indicating that the learned user profiles better capture actual user behavior patterns. Detailed experimental settings and results are reported in Appendix~\ref{app:rq3}.

\subsection{Case Study}
To further illustrate the structure and characteristics of the news summary, the domain-aware user profiles, and user commenting feature, we present a representative case study in Appendix~\ref{app:case_study}.

%% file: tables/general_performance.tex
\begin{table*}
  \setlength{\tabcolsep}{2.4mm}{
  \caption{Comparison with single-domain and cross-domain FND baselines in the general setting (\%).}
  \label{tab:general}
  \small
  \begin{threeparttable}
  \begin{tabular}{l|cccc|cccc|cccc}
  \hline \hline 
  \multirow{2}{*}{ Methods } & \multicolumn{4}{c|}{ Polit. + Enter. + Covid. $\rightarrow$ Polit. } & \multicolumn{4}{c|}{ Polit. + Enter. + Covid. $\rightarrow$ Enter. }  & \multicolumn{4}{c}{ Polit. + Enter. + Covid. $\rightarrow$ Covid. }\\
  \cline{2-13}
  & Prec. & Rec. & F1 & AUC & Prec. & Rec. & F1 & AUC & Prec. & Rec. & F1 & AUC \\
  \hline
BERT & $86.25 $ & $86.51 $ & $85.62 $ & $93.59 $ & $51.41 $ & $50.70 $ & $50.84 $ & $61.35 $ 
& $52.82 $ & $52.51 $ & $52.66 $ & $65.28 $ \\
dEFEND & $88.33 $ & $77.94 $ & $82.81 $ & $83.42 $ & $56.51 $ & $63.71 $ & $59.90 $ & $75.11 $ 
& $59.24 $ & $57.69 $ & $58.45 $ & $75.90 $ \\
FANG & $61.19 $ & $62.02 $ & $61.83 $ & $79.18 $ & $51.58 $ & $52.12 $ & $50.50 $ & $62.59 $ 
& $53.66 $ & $53.14 $ & $53.40 $ & $64.20 $ \\
DUCK & $84.90 $ & $77.46 $ & $78.63 $ & $85.40 $ & $74.77 $ & $67.31 $ & $70.11 $ & $75.60 $ &
 $75.63 $ & $71.26 $ & $73.38 $ & $77.61 $ \\
\hline
GPT-4o-mini & $67.59 $ & $66.11 $ & $63.84 $ & $66.11 $ & $57.32 $ & $58.28 $ & $43.13 $ & $58.28 $ 
 & $60.74 $ & $62.64 $ & $61.68 $ & $60.15 $ \\
ARG & $88.00 $ & $88.47 $ & $88.06 $ & $92.57 $ & $75.08 $ & $73.81 $ & $74.08 $ & $86.39 $ 
 & $81.80 $ & $80.38 $ & $81.08 $ & $89.80 $ \\
DELL & $90.82 $ & $90.40 $ & $90.61 $ & $95.41 $ & $78.61 $ & $79.54 $ & $79.07 $ & $87.98 $ 
 & $81.32 $ & $80.95 $ & $81.13 $ & $90.39 $ \\
\hline
EANN & $61.04 $ & $67.83 $ & $64.26 $ & $66.76 $ & $66.30 $ & $70.03 $ & $68.11 $ & $70.81 $
& $65.91 $ & $66.12 $ & $66.01 $ & $68.22 $ \\
REAL-FND & $78.57 $ & $82.43 $ & $80.45 $ & $85.28 $ & $72.83 $ & $75.79 $ & $74.28 $ & $78.58 $ 
& $69.57 $ & $70.45 $ & $70.01 $ & $75.33 $\\
EDD & $85.09 $ & $87.30 $ & $86.18 $ & $93.25 $ & $81.88 $ & $83.64 $ & $82.75 $ & $91.84 $ 
& $82.56 $ & $83.23 $ & $82.89 $ & $93.97 $ \\
M3FEND & $92.50 $ & $91.33 $ & $91.69 $ & $93.72 $ & $81.19 $ & $81.82 $ & $81.50 $ & $90.49 $
& $83.02 $ & $82.58 $ & $82.80 $ & $91.89 $\\
UPDATE & $92.16 $ & $91.13 $ & $91.64 $ & $92.05 $ & $81.91 $ & $81.89 $ & $81.90 $ & $88.08 $ 
 & $85.20 $ & $83.62 $ & $84.40 $ & $89.94 $ \\
MMHT & $\underline{94.97}\tnote{2} $ & $\underline{95.25} $ & $\underline{95.11} $ & $\underline{97.68} $ & $\underline{88.01} $ & $\underline{85.74} $ & $\underline{86.86} $ & $\underline{93.16} $ 
 & $\underline{90.34} $ & $\underline{87.36} $ & $\underline{88.83} $ & $\underline{95.26} $ \\
\hline
\textbf{DAUD} & $\mathbf{96.21}\tnote{1} $ & $\mathbf{96.78} $ & $\mathbf{96.49} $ & $\mathbf{98.16} $ & $\mathbf{90.11} $ & $\mathbf{88.14} $ & $\mathbf{89.11} $ & $\mathbf{95.20} $ 
 & $\mathbf{93.67} $ & $\mathbf{90.90} $ & $\mathbf{92.26} $ & $\mathbf{96.20} $ \\
  \hline 
\multicolumn{1}{c|}{\textit{Improvement}\tnote{3}} & $+1.31\%$ & $+1.61\% $ & $+1.45\% $ & $+0.49\% $ & $+2.39\%$ & $+2.80\% $ & $+2.59\% $ & $+2.19\% $ & $+3.69\%$ & $+4.05\% $ & $+3.86\% $ & $+0.99\% $\\
  \hline \hline
  \end{tabular}

  \begin{tablenotes}
  \footnotesize
  \item[1] A value in bold font indicates the best performance. 
  \item[2] An underlined value indicates the performance of best-performing baseline approaches.
  \item[3] The improvement of DAUD over the best-performing baseline approaches, and the improvement is significant at $p<0.05$.
  \end{tablenotes}

  \end{threeparttable}
  }
\end{table*}

%% file: tables/unseen_performance.tex
\begin{table*}
  \setlength{\tabcolsep}{2.4mm}{
  \caption{Comparison with CD-FND baselines in the unseen-domain setting (\%).}
  \label{tab:unseen}
  \small
  \begin{threeparttable}
  \begin{tabular}{l|cccc|cccc|cccc}
  \hline \hline 
  \multirow{2}{*}{ Methods } & \multicolumn{4}{c|}{ Enter. + Covid. $\rightarrow$ Polit. } & \multicolumn{4}{c|}{ Polit. + Covid. $\rightarrow$ Enter. }  & \multicolumn{4}{c}{ Polit. + Enter. $\rightarrow$ Covid. }\\
  \cline{2-13}
  & Prec. & Rec. & F1 & AUC & Prec. & Rec. & F1 & AUC & Prec. & Rec. & F1 & AUC \\
\hline
EANN & $54.52 $ & $50.25 $ & $52.30 $ & $53.70 $ & $58.14 $ & $57.79 $ & $57.96 $ & $60.94 $  & $55.38 $ & $54.71 $ & $55.04 $ & $56.83 $ \\
REAL-FND & $\underline{76.74}\tnote{2} $ & $75.15 $ & $\underline{75.94} $ & $77.38 $ & $67.96 $ & $65.49 $ & $66.70 $ & $66.62 $ & $69.02 $ & $68.49 $ & $68.75 $ & $\underline{71.15} $\\
EDD & $73.21 $ & $76.24 $ & $74.69 $ & $75.10 $ & $58.26 $ & $59.18 $ & $58.72$ & $63.97 $ & $64.28 $ & $65.75 $ & $65.01 $ & $64.37 $ \\
M3FEND & $65.43 $ & $64.11 $ & $64.76 $ & $65.44 $ & $66.30 $ & $61.53 $ & $63.83 $ & $64.40 $ & $63.56 $ & $64.71 $ & $64.13 $ & $66.83 $\\
UPDATE & $55.28 $ & $52.33 $ & $53.76 $ & $55.90 $ & $60.17 $ & $62.55 $ & $61.34 $ & $65.68 $ & $59.35 $ & $58.14 $ & $58.74 $ & $61.06 $ \\
MMHT & $72.06 $ & $\underline{76.31} $ & $74.12 $ & $\underline{78.14} $ & $\underline{71.27} $ & $\underline{72.18} $ & $\underline{71.72} $ & $\underline{73.06} $ & $\underline{69.23} $ & $\underline{72.00} $ & $\underline{70.59} $ & $70.17 $\\
\hline
\textbf{DAUD} & $\mathbf{77.55}\tnote{1} $ & $\mathbf{78.42} $ & $\mathbf{77.98} $ & $\mathbf{79.16} $ & $\mathbf{73.01} $ & $\mathbf{73.52} $ & $\mathbf{73.26} $ & $\mathbf{75.44} $ & $\mathbf{72.08} $ & $\mathbf{73.98} $ & $\mathbf{73.02} $ & $\mathbf{74.12} $\\
  \hline 
\multicolumn{1}{c|}{\textit{Improvement}\tnote{3}} & $+1.06\%$ & $+2.77\% $ & $+2.69\% $ & $+1.31\% $ & $+2.44\%$ & $+1.86\% $ & $+2.15\% $ & $+3.26\% $ & $+4.12\%$ & $+2.75\% $ & $+3.44\% $ & $+4.17\% $\\
  \hline \hline
  \end{tabular}

  \begin{tablenotes}
  \footnotesize
  \item[1] A value in bold font indicates the best performance. 
  \item[2] An underlined value indicates the performance of best-performing baseline approaches.
  \item[3] The improvement of DAUD over the best-performing baseline approaches, and the improvement is significant at $p<0.05$.
  \end{tablenotes}

  \end{threeparttable}
  }
\end{table*}

%% file: figures/ablation_study.tex
\begin{figure}
  \centering
  \includegraphics[scale=0.4]{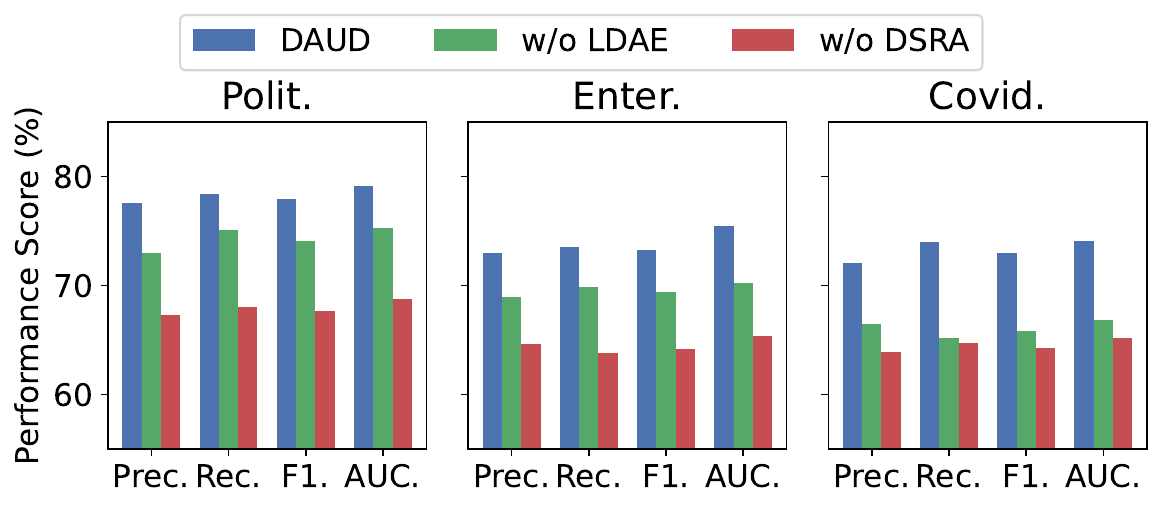}
  \caption{Ablation Study of DAUD in unseen-domain CD-FND setting.}
  \label{fig:ablation}
  \Description{ablation study}
\end{figure}

%% file: sections/5conclusion.tex
\section{Conclusions and Future Work}
In this work, we tackle the challenges of CD-FND, with a particular focus on improving performance in the unseen-domain setting.
We propose DAUD, a novel framework that integrates an LLM-based module that enriches news and user features with high-level semantics; and a relation-aware alignment module that captures hierarchical contextual relations and learns domain-shared representations.
Extensive experiments in three domains demonstrate that our approach consistently outperforms existing state-of-the-art baseline methods, showing clear improvements in both general and unseen-domain CD-FND settings.
For future work, we plan to explore DAUD’s extension to scenarios without common users by improving the effectiveness of domain-shared feature learning.

%% file: sections/6appendix.tex
\section{Prompt Design}
\label{app:prompts}
We present the News Feature Extraction Prompt and the Comment Generation Prompt used in our methodology.
\begin{promptblue}[News Feature Extraction Prompt]
Recently, the user browsed a news article, its article content is: \{\,\textcolor{blue}{\textit{news article}}\,\}.

\textbf{Your task is to analyze the content of the news text and then summarize the characteristics of true and fake information within the news.}

\textbf{Follow these steps: }

(1) Identify and ignore website noise, such as ads, image captions, or unrelated links.

(2) Analyze the news text based on: News Domain, Sentiment (e.g., neutral, emotional, exaggerated), Structural Features, Logical Consistency, Source Credibility (Check sources of quotes and be suspicious).
\end{promptblue}

\begin{promptlightgreen}[Comment Generation Prompt]
Recently, the user has posted multiple comments on different news articles, the comments are listed as follows: \{\,\textcolor{blue}{\textit{comments list}}\,\}. \textbf{Your task is to analyze these comments and summarize the user’s typical tone and commenting style.}

\textbf{Follow these steps:}

(1) Identify and ignore parts of the comments that are formulaic (e.g., hashtags, URLs, emojis, or repost tags) unless they contribute meaningfully to the tone.

(2) Analyze the comments based on: Tone (e.g., sarcastic, sincere, angry, humorous, accusatory, supportive); Intent (e.g., inform, provoke, agree, debunk, mock); Linguistic Style (e.g., formal, casual, concise, emotionally charged, rhetorical); Stance Consistency (e.g., does the user consistently take a certain side or shift based on domain or content); Targeting Pattern (e.g., does the user address individuals, institutions, abstract ideas, or the public).
\end{promptlightgreen}

\section{Datasets}
\label{app:datasets}

Table~\ref{tab:characteristics} summarizes statistics of datasets of Politics, Entertainment and Covid-19.
These domains exhibit substantial lexical and semantic differences, which pose a significant challenge for CD-FND in the unseen-domain setting. For example, as shown in the word clouds in Fig.~\ref{fig:word_cloud}, political news frequently contains terms related to words like policy and election; entertainment news is dominated by words like celebrity and idol; whereas COVID-19 news is characterized by words like pandemic and outbreak.
\section{Baseline Methods}
\label{app:baseline}
\textbf{Single-domain traditional methods.} 
We select four representative traditional methods trained and evaluated within a single domain: 
(1) \textbf{BERT}~\cite{devlin2018bert}, a pre-trained transformer encoder fine-tuned for fake news classification; (2) \textbf{dEFEND}~\cite{shu2019defend}, which models engagements between news articles and user comments; (3) \textbf{FANG}~\cite{nguyen2020fang}, which learns graph-based social context representations from user–news interactions; and (4) \textbf{DUCK}~\cite{tian2022duck}, which aggregates multiple graph-derived engagement features to improve veracity prediction. 
\textbf{Single-domain LLM-based methods.} 
We further select three representative LLM-based methods that apply large language models to fake news detection within a single domain: 
(1) \textbf{GPT-4o-mini}, a lightweight variant of GPT-4o that performs direct veracity prediction using its generative reasoning abilities; (2) \textbf{ARG}~\cite{hu2024bad}, which distills knowledge from a large language model into a smaller task-specific model for news classification; and (3) \textbf{DELL}~\cite{wan2024dell}, which leverages LLM-generated user reactions and explanations, combining them with a GNN backbone to perform fake news detection.
\textbf{Cross-domain traditional methods.}
We compare DAUD with six representative cross-domain FND models: 
(1) \textbf{EANN}~\cite{wang2018eann}, an adversarial learning framework that extracts domain-shared features from news content; (2) \textbf{Real-FND}~\cite{mosallanezhad2022domain}, a reinforcement-learning-based model that captures domain-shared information from both news content and user engagements; (3) \textbf{EDD}~\cite{silva2021embracing}, which separates domain-shared and domain-specific features through multi-level adversarial training; (4) \textbf{M3FEND}~\cite{zhu2022memory}, which models domain discrepancies by leveraging multiple types of textual features derived from news content; (5) \textbf{UPDATE}~\cite{yang2024update}, which performs cross-domain knowledge transfer by integrating user engagement signals with news content; (6) \textbf{MMHT}~\cite{yang2025macro}, which employs a hierarchical disentangler to achieve more effective cross-domain knowledge transfer.
Note that we do not include the LLM-based CD-FND methods introduced in the related work as baselines, as they do not explicitly transfer domain-shared knowledge across domains. Because our experiments are designed to evaluate the effectiveness of cross-domain knowledge transfer, these methods are not directly comparable within our evaluation settings.

\input{tables/data}
\input{figures/word_cloud}

\section{Implementation Details}
\label{app:implement}
In LDAE, we use the API of GPT-4o-mini to generate enriched news features, and user features as well as augmented user engagements. In DSRA, we use a RoBERTa encoder to obtain embeddings for news content, news summaries, user profiles, historical news, and comments. We set the maximum input length to 256 and use a hidden size of 768 for all RoBERTa-based embeddings. For training, we use the AdamW optimizer with a learning rate of 1e-4, a linear warm-up ratio of 10\%, and a batch size of 32. The model is trained for 100 epochs with a dropout rate of 0.1. We report four commonly used metrics for CD-FND: Precision (Prec.), Recall (Rec.), F1-score (F1), and AUC, and present the average performance across five independent runs.

\section{Experimental Details for RQ3}
\label{app:rq3}

\input{figures/comment_rating}
Firstly, we evaluate the quality of the comments generated by LDAE using an LLM~\cite{chiang2023can, kim2023prometheus}. We randomly sample 500 generated comments across all domains and prompt the LLM with two questions: (1) "Does  the user’s comment on the news match the user profile?" and (2) "Does the comment relate to the news?" The LLM provides a score for each question on a five-point Likert scale~\cite{allen2007likert}. For clarity, we map the numeric scale (1–5) to qualitative labels (from “very unlikely” to “very likely”). The final rating for each comment is obtained by averaging the scores of the above two questions. For comparison, we evaluate an equal number of real comments using the same procedure. As shown in Fig.~\ref{fig:comment_rating}, the averaged scores of the two evaluation questions indicate that the generated comments are better than the real ones overall, and their variances are consistently smaller across all domains, suggesting that the generated comments are more stable in quality.

\input{tables/engagement_prediction}
Secondly, to examine whether the learned user profiles encode real user preferences, we conduct an auxiliary engagement prediction task. This task takes the user and news representations as input, and uses a prompt-based module to predict whether the user would engage with the news. We use user engagement records in the dataset as ground-truth labels for evaluation, and report Precision, Recall, Accuracy, Micro F1, and Macro F1. For comparison, we replace our learned user profiles with the predefined user attributes used in~\cite{nan2024let}, and feed these attributes into our engagement prediction prompt for prediction and evaluation. As shown in Table~\ref{tab:engagement_prediction}, LDAE’s domain-aware user agent achieves substantially stronger performance than this attribute-based variant, indicating that our learned user profiles better capture actual user behavior patterns than predefined attributes, thereby providing more informative representations for CD-FND.

\input{tables/case_study}
\section{Case Study}
\label{app:case_study}
Table~\ref{tab:case_study} presents two representative cases. Case~1 shows the news summary extracted from a political fake news article using the News Feature Extraction Prompt, together with the intermediate user profile generated by LDAE based on the historical engagement with this article through the User Feature Update Prompt. Case~2 presents the commenting features obtained using the Comment Generation Prompt, extracted from five real comments written by the user.

Beyond simply displaying example outputs, these cases also demonstrate the analytical richness enabled by our prompt designs. The News Feature Extraction Prompt captures not only high-level domain information but also fine-grained news features, including sentiment, narrative structure, logical consistency, and source credibility. Similarly, the User Feature Update Prompt produces user profiles that reveal coherent behavioral patterns, engagement preferences, and stance tendencies across domains. Meanwhile, the Comment Generation Prompt distills commenting behavior into concise yet semantically meaningful features, highlighting linguistic style, and emotional tone.

%% file: tables/data.tex
\begin{table}
  \setlength{\tabcolsep}{1.5mm}{
  \caption{The characteristics of three experimental datasets.}
  \label{tab:characteristics}
  \small
  \begin{tabular}{lccc}
  \hline \hline {Statistics} & Polit. & Enter. & Covid. \\
  \hline 
  \# True News & 339 & 13,194 & 4,750 \\
  \# Fake News & 347 & 3,692 & 1,317 \\
  \hline 
  \# Users & 174,110 & 82,871 & 949\\ 
  \# User Engagements & 419,358 & 225,467 & 16,955\\
  \hline \hline
  \end{tabular}}
\end{table}

%% file: figures/word_cloud.tex
\begin{figure}[t]
    \centering
    \begin{subfigure}[b]{0.15\textwidth}
        \centering
        \includegraphics[width=\textwidth]{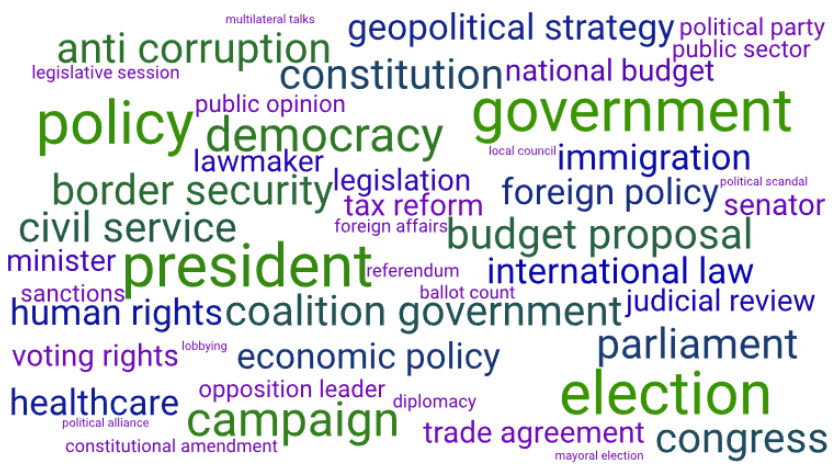}
        \caption{Politics}
        \label{fig:a}
    \end{subfigure}
    \hspace{1mm}
    \begin{subfigure}[b]{0.13\textwidth}
        \centering
        \includegraphics[width=\textwidth]{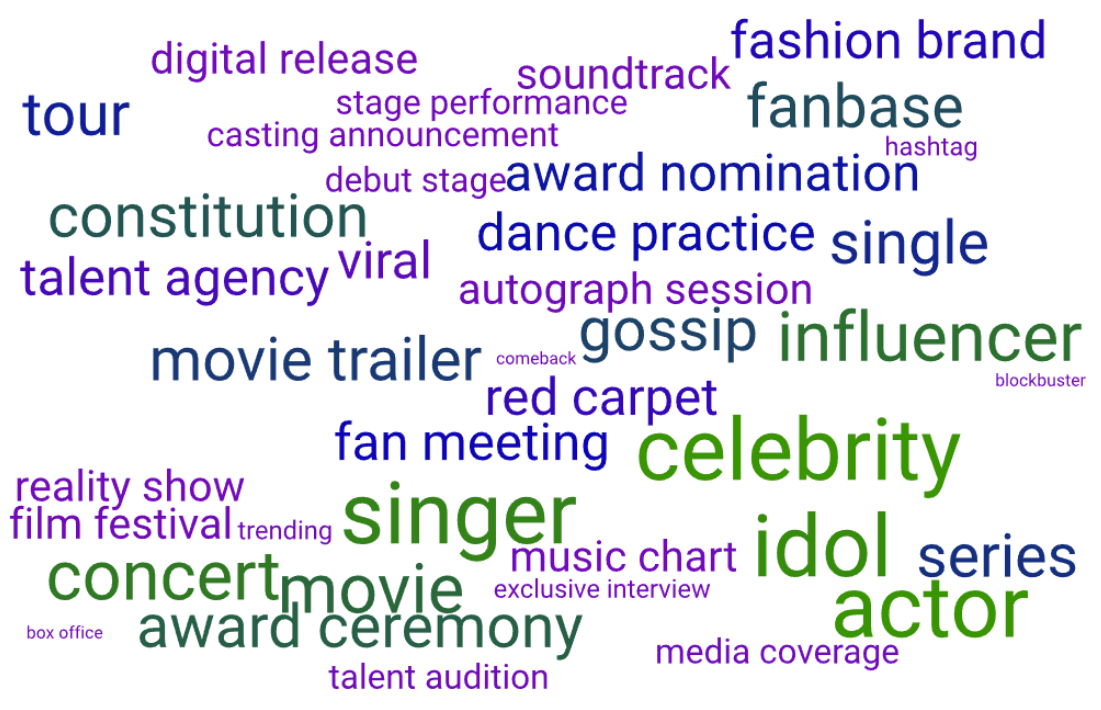}
        \caption{Entertainment}
        \label{fig:b}
    \end{subfigure}
    \hspace{1mm}
    \begin{subfigure}[b]{0.14\textwidth}
        \centering
        \includegraphics[width=\textwidth]{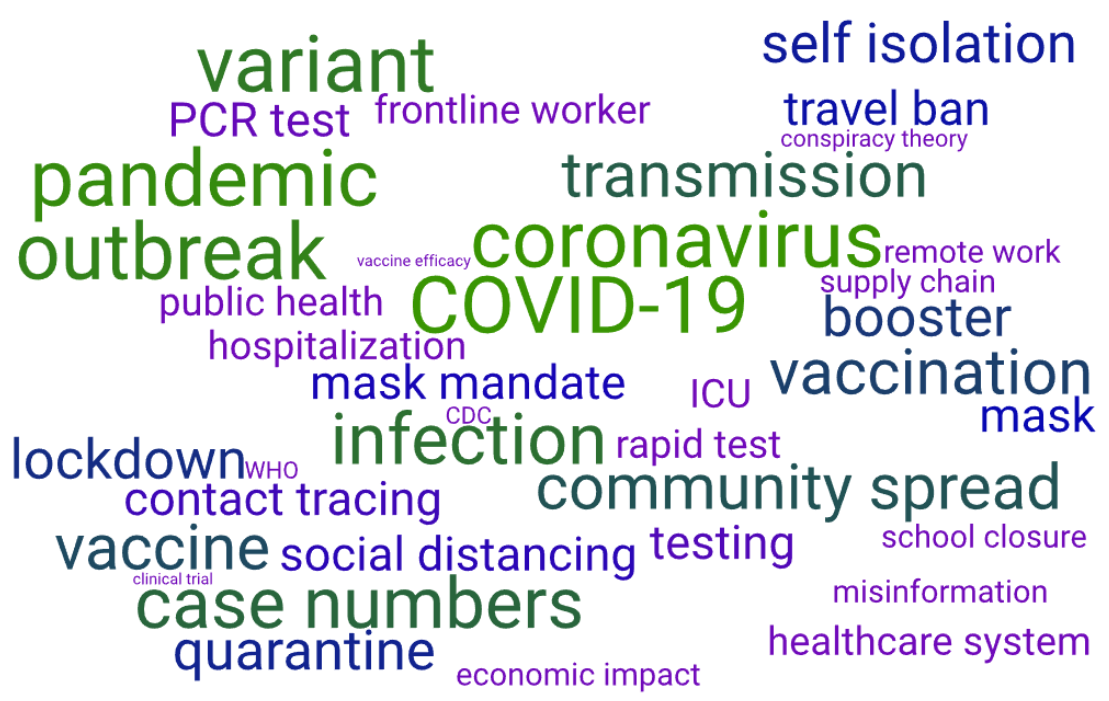}
        \caption{COVID-19}
        \label{fig:c}
    \end{subfigure}
    \caption{Word Clouds for Politics, Entertainment, and COVID-19 domains.}
    \label{fig:word_cloud}
\end{figure}

%% file: figures/comment_rating.tex
\begin{figure}
  \centering
  \includegraphics[scale=0.43]{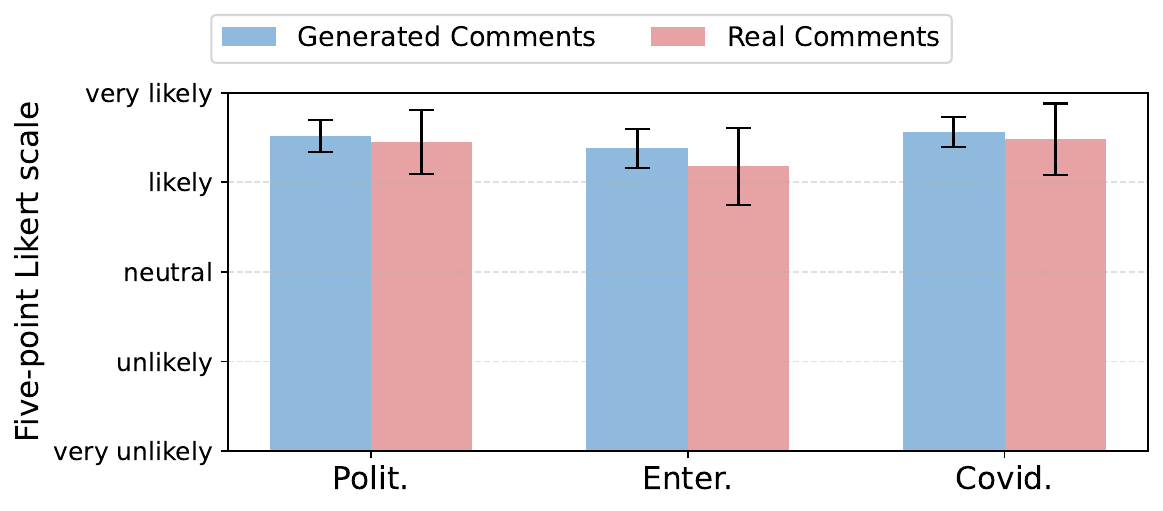}
  \caption{LLM-based Evaluation of Generated and Real Comments.}
  \label{fig:comment_rating}
  \Description{comment rating}
\end{figure}

%% file: tables/engagement_prediction.tex
\begin{table}[t]
\centering
\caption{Engagement prediction results of the domain-aware user agent.}
\begin{tabular}{l|ccccc}
\hline \hline
Metric & Prec. & Rec. & Acc. & Micro F1 & Macro F1 \\
\hline
Ours & \textbf{1.00} & \textbf{0.76} & \textbf{0.76} & \textbf{0.76} & \textbf{0.86} \\
Baseline & 0.51 & 0.22 & 0.40 & 0.31 & 0.33 \\
\hline \hline
\end{tabular}
\label{tab:engagement_prediction}
\end{table}

%% file: tables/case_study.tex
\begin{table*}[t]
\centering
\small
\caption{Case studies of news summary and user profile as well as user commenting features.}
\label{tab:case_study}
\begin{tabular}{p{0.97\linewidth}}
\toprule

\textbf{Case 1: News Summary and User Profile} \\
\hline
\textbf{Generated News Summary (fake news):} 

This article belongs to the \ul{\textbf{political domain}} and contains \ul{\textbf{highly exaggerated and satirical sentiment}}. It \ul{\textbf{lacks a coherent news structure and includes implausible claims}} such as mass mind control via a "Russian space-beam" and fabricated legal rulings criminalizing Trump support. Though real names like Mike Rogers and Dmitry Peskov are mentioned, \ul{\textbf{their quotes are clearly fictionalized and not verifiable through credible sources}}. The content relies on \ul{\textbf{absurd humor and parody}}. There are \ul{\textbf{no official links or evidence}} backing any major claims, and the use of stereotypes and conspiracy language undermines any factual plausibility. \\[2pt]

\textbf{LDAE-Inferred User Profile:} 

I am a \ul{\textbf{politically}} engaged user who often interacts with \ul{\textbf{exaggerated, conspiratorial, or satirical content}}, especially when it targets controversial political figures or polarizing issues in the United States. I tend to repost news that \ul{\textbf{dramatizes or ridicules establishment narratives}}, even when the claims are implausible or absurd, as long as they align with anti-mainstream or anti-Trump sentiment presented in an entertaining or provocative manner. I dislike \ul{\textbf{dry or neutral reporting}}, strictly evidence-based political analysis, or any content that appears to defend or rationalize Trump or his supporters. \\[2pt]

\midrule

\textbf{Case 2: User Commenting Features} \\
\hline
\textbf{Real User Comments:}
\begin{itemize}[leftmargin=10pt]
    \item[1)] DID YOU KNOW? Car crashes are the leading cause of death for teens ages 16 to 19, according to the Centers for Dise… https://t.co/I1R2XlDqkd;  
    \item[2)] ALERT: The Centers for Disease Control and Prevention released a statement Friday reminding Americans to not eat ro… https://t.co/hc7UkHHAQo;  
    \item[3)] House Budget Bill Includes Welfare Reform: Included in a \$33 billion spending bill passed by Michigan's House on... http://bit.ly/iQrZa6;  
    \item[4)] Looking for something to watch this Easter Sunday? NBC's next live musical spectacular, the groundbreaking rock ope… https://t.co/gHb8vXCE5l; 
    \item[5)] April the Giraffe is pregnant again https://t.co/dh3EoNDjHH https://t.co/sCtPIgZ5eY;
\end{itemize}

\textbf{Generated Commenting Features:} 

The comment style of this user is primarily \ul{\textbf{informative and alert-driven}}, with a tone that is straightforward and occasionally urgent but not emotionally charged. The user favors \ul{\textbf{concise, headline-like statements}} that highlight important or noteworthy facts. The linguistic style is \ul{\textbf{formal to semi-formal, employing clear, declarative sentences without slang or rhetorical flourish}}, suggesting an emphasis on clarity over personality or humor. The user’s stance is \ul{\textbf{consistently neutral and fact-focused}}, avoiding overt opinion or bias, and tends to address broad public audiences rather than specific individuals or entities, aiming to disseminate news or updates from authoritative sources like the CDC or legislative bodies. Overall, the user’s commenting pattern resembles a \ul{\textbf{news-curation style, emphasizing awareness and public safety, with little variation in tone or target across diverse topics}}. \\[2pt]

\bottomrule
\end{tabular}
\end{table*}